\documentclass[doublecol]{epl2} 
\usepackage{epstopdf}

\usepackage{graphicx}
\usepackage{amsfonts}
\usepackage{amsmath}
\usepackage{mathtools}
\usepackage{float}
\usepackage{epsfig}
\usepackage[dvipsnames]{xcolor}
\usepackage{etoolbox}
\usepackage{gensymb}
\usepackage{multirow}
\usepackage{multicol}
\usepackage{subcaption}
\usepackage[mathscr]{euscript}
\usepackage{pifont}
\usepackage{hyperref}
\hypersetup{colorlinks=true, citecolor=black, linkcolor=black}
\usepackage{physics}
\newcommand{\hot}{T_{\mathrm{I}}}
\newcommand{\cold}{T_{\mathrm{F}}}

\newcommand{\maxcoldSS}{0.0080}

\begin{document}

\title{Building a ``trap model'' of glassy dynamics from a local structural predictor of rearrangements}

\author{S. A. Ridout\inst{1,2} \and I. Tah\inst{1,3} \and A. J. Liu\inst{1,4}}
\shortauthor{S. A. Ridout \etal}

\institute{                    
  \inst{1}Department of Physics and Astronomy, University of Pennsylvania -- Philadelphia, PA 19104, USA
  
  \inst{2}Department of Physics, Emory University -- Atlanta, GA 30322, USA
  
  \inst{3} Speciality Glass Division, CSIR-Central Glass and Ceramic Research Institute -- Kolkata 700032, India

  \inst{4} Center for Computational Biology, Flatiron Institute, Simons Foundation -- New York, NY 10010, USA

}

\abstract{
Here we introduce a variation of the trap model of glasses based on softness, a local structural variable identified by machine learning, in supercooled liquids.  Softness is a particle-based quantity that reflects the local structural environment of a particle and characterizes the energy barrier for the particle to rearrange.  As in the trap model, we treat each particle's softness, and hence energy barrier, as evolving independently.  We show that such a model reproduces many qualitative features of softness, and therefore makes qualitatively reasonable predictions of behaviors such as the dependence of fragility on density in a model supercooled liquid. We also show failures of this simple model, indicating features of the dynamics of softness that may only be explained by correlations.}

\maketitle

\section{Introduction}

The connection between dynamics and microscopic structure in supercooled liquids has bedeviled the glass transition field for many decades. Recently, however, the field has reached a consensus that structural heterogeneities are important to the dynamics of supercooled liquids\cite{ediger2000review,tah2021understanding}.  Isoconfigurational ensemble simulations give direct evidence of heterogeneous mobility, which must be structural in origin~\cite{widmer2004reproducible,widmer2008irreversible}. More recently, machine-learning methods have identified structural variables that correlate strongly with dynamics ~\cite{cubuk2015identifying,schoenholz2016structural,schoenholz2017relationship,sussman2017,tah2021,tah2022fragility,ridout2022,ma2019heterogeneous,cubuk2020,yang2022,bapst2020unveiling,shiba2023,jung2022predicting,boattini2021averaging,boattini2020autonomously}.  

Before such microscopic observations, there was already a consensus that heterogeneity in dynamics is important, allowing for the explanation of e.g. nonexponential relaxation and breakdown of the Stokes-Einstein relation~\cite{cavagna2009supercooled,cicerone1996enhanced}.   One particularly simple---and therefore appealing---phenomenological model of heterogeneous dynamics is the trap model \cite{monthus1996models,bouchaud1992weak}.  In this model, a coarse-grained region (or a single particle) is presumed to have some distribution of possible ``trap'' energies $\varrho{\left(E\right)}$. The model assumes that the system ``escapes'' these traps with an Arrhenius rate characterized by an energy barrier $E$, landing in a new trap selected without memory from a temperature-independent distribution $\varrho{\left(E\right)}$. Although this model neglects spatial correlations in these energy barriers, it accounts qualitatively for aging and stretched-exponential relaxation. It has served as a useful starting point for powerful models of glass phenomenology, such as the Soft Glassy Rheology model \cite{sollich1997} as well as models of facilitated thermal relaxation \cite{Rehwald2010,Rehwald2012,guiselin2022}.

The trap model assumes that particles escape their traps with a rate $e^{-\beta E}$, as if escaping from a trap always requires crossing a barrier at $E=0$.  This assumption connects the dynamics to the energy $E$ and its distribution. There is a natural connection between the trap model and a particular machine-learned structural variable, called softness, $S$. This connection arises because the probability that a particle of a given softness will rearrange, $P(R|S)$, has been shown in a wide variety of systems~\cite{schoenholz2016structural,schoenholz2017relationship,sharp2018,sussman2017,cubuk2020,tah2021,tah2022fragility,yang2022} to have an Arrhenius temperature dependence. This result implies that the softness of a particle tells us the typical energy barrier, $\Delta E(S)$, that the particle surmounts in order to rearrange. Thus, we can construct a trap model based on softness, in which we replace the barriers $E$ in the original trap model with $\Delta E(S)$, and introduce an underlying softness distribution $\varrho{\left(S\right)}$ that replaces $\varrho{\left(E\right)}$ in the trap model. 

Any local structural variable that is correlated strongly enough with the dynamics can be used in place of softness in the analysis that follows. We focus on softness because it has already been demonstrated that $P(R|S)$ is Arrhenius~\cite{schoenholz2016structural,schoenholz2017relationship,sharp2018,sussman2017,cubuk2020,tah2021,tah2022fragility,yang2022}, but any reasonable local predictor should have this property. Note that other approaches to studying the dynamics and equilibrium distributions of particular local structures have been considered~\cite{lerner2009statistical,robinson2019morphometric}. 

Here we develop the simplest possible model of the dynamics of $S$, namely a trap-like model in which each particle's softness (or equivalently, energy barrier) evolves independently.  In spite of the obvious oversimplification involved, we find that this model produces many reasonable predictions.  It provides a simple explanation of how $\langle S\rangle$ should decrease with temperature and thus how we may see a larger energy barrier with decreasing $T$, producing a super-Arrhenius dependence of relaxation time $\tau$ on temperature. In a class of model glass-forming liquids, we find that we are able to account for much of the difference in fragility between different systems using this simple model. We also show the limits of the trap-like model, showing how it fails to reproduce some important qualitative features of dynamical heterogeneity and aging.

\section{Simulation models and methods}

The softness $S_i$ is a linear combination of structural variables that describe the local structure near particle $i$. Similarly to \cite{cubuk2015identifying}, we define $M=166$ (or $266$, depending on system) structural variables $g_{\alpha, i}$, computed using inherent-structure (IS) positions, that describe the local structure surrounding particle $i$ through both density and bond-angle information (details in supplement).  The softness is then

\begin{equation}
S_i = w_0 + \sum_{\alpha} w_\alpha g_{\alpha, i} ,
\label{eq:Sdef}
\end{equation}
where $\mathbf{w}$ is chosen (``trained'') so that $S_i$ is maximally predictive of whether or not particle $i$ is rearranging.

We study two model supercooled liquids in 3D, for which softness has previously been shown to correlate with rearrangements.  
The first is the standard Kob-Andersen Lennard-Jones (KA) system, a mixture of 80\% large and 20\% small particles~\cite{kobandersen}. We study systems of $N=10000$ particles. Our analysis follows \cite{schoenholz2016structural}: we train $w_\alpha$ using data from $T=0.47$ and apply it at all temperatures studied, focusing only on large particles, and training $w_\alpha$ so that $S_i$ predicts the rearrangement indicator $p_{\mathrm{hop}}$. At low $T$, some systems crystallize and are excluded from analysis.

The second system (abbreviated as SS for soft sphere) is a $50:50$ bidisperse liquid of of $N=10976$ particles with harmonic repulsion, as studied in \cite{berthier2009compressing,berthier2009glass,tah2022kinetic,tah2022fragility}.  The fragility of the model is tuned by adjusting the density. As in \cite{tah2022fragility}, a single vector $w_\alpha$ has been trained to define softness at all temperatures $T$ and number densities $\rho$ studied.  We analyze softness for the small particles. We train $S$ to predict the rearrangement indicator $D^2_{\mathrm{min}}$ (data from ~\cite{tah2022fragility}), but unlike ~\cite{tah2022fragility}, compute $g_\alpha$ using IS positions.  

Note that rearrangement probabilities must be multiplied by $T$-independent factors to convert them to a rate of rearrangement per unit time (details in supplement).

Here, it is important that softness has a single definition in each system, instead of a different definition at each temperature and density. This allows us to predict properties at one $T$ from observations at another $T$.

As in earlier analyses~\cite{schoenholz2016structural,sharp2018,schoenholz2017relationship,sussman2017,cubuk2020,tah2021,tah2022fragility,yang2022,ridout2022,ma2019heterogeneous}, the probability for a particle to rearrange within a time interval $\tau_R$ at temperature $T$, $P{\left(R|S,T\right)}$, is found to depend roughly exponentially on $S$ (\cite{schoenholz2016structural} fig. 2(a), \cite{tah2022fragility} fig. 1.), i.e.

\begin{equation}
    P{\left(R|S,T\right)} \approx A{\left(T\right)} e^{\gamma{\left(T\right)} S}. 
\end{equation}

Thus, $S$ reveals heterogeneity in rearrangement rates at a single-particle level. For particles of a given $S$, the rearrangement rate appears Arrhenius,  

\begin{equation}
    P{\left(R|S,T\right)} \approx    e^{\Sigma{\left(S\right)} - \Delta E{\left(S\right)} / T},
\end{equation}
where $\Delta E{\left(S\right)}$ and $\Sigma{\left(S\right)}$ are temperature-independent factors which we thus interpret as an energy barrier to rearrangement and an entropic contribution to the rate of rearrangement for a given softness (~\cite{schoenholz2016structural} fig. 2, \cite{tah2022fragility} fig. 1).
The approximate exponential behaviour of $P{\left(R|S\right)}$ corresponds to approximate linearity of $\Sigma{\left(S\right)}$ and $\Delta E{\left(S\right)}$. Thus, $\Delta E{\left(S\right)}$, inferred from observations of rearrangements in a simulation or particle-resolved experiment, realizes precisely the connection between state and dynamics that is assumed for $E$ in the trap model.

\section{The trap-like model}

Consider a model of $M$ sites, each representing $m$ particles, so that $N = m M$. Each site $i$ has a softness $S_i$. In the end we will take $m=1$ since our machine-learned softness is defined for single particles.

Rearrangement of single particles is observed to have a rate which is Arrhenius, with an energy barrier roughly linear in $S$, as discussed above.  Thus, the simplest possible dynamical model is spatially-resolved model of independent traps.  We postulate the existence of a density $\varrho{\left(S\right)}$ of states for $S$; since $P{\left(R|S\right)}$ becomes independent of $S$ above $T_0$~\cite{schoenholz2016structural}, we will find that $\varrho{\left(S\right)} = P{\left(S|T \geq T_0\right)}$.  

A site is assumed to ``rearrange'' with rate 
\begin{equation}
k{\left(S, T\right)} \equiv \frac{1}{\tau_R} P{\left(R |S,T\right)} =  e^{\Sigma{\left(S\right)} - \Delta E{\left(S\right)} /T}, \label{eq:rate}
\end{equation} 
Here $\tau_R$ is the window of time used to measure P${\left(R |S, T\right)}$. When a particle rearranges, we assume that it forgets its history, and its new softness $S'$ is drawn from $\varrho{\left(S'\right)}$.

As stated above above this model is formally equivalent to a trap model; e.g.if $\Delta E$ and $\Sigma$ are both perfectly linear in $S$, this is a trap model with a Gaussian distribution of energy barriers.  However, $S$ need not be directly related to the energy barrier $\Delta E$. 
Note that unlike the standard trap model\cite{bouchaud1992weak}, or a model of Gaussian traps in which the energy variance scales with $N$ as in the random energy model \cite{Cammarota2018,arous2002aging}, this model contains no sharp transition.~\cite{monthus1996models}

$P{\left(S\right)}$ then obeys the master equation

\begin{equation}
   \dv{P_t{\left(S\right)}}{t} = - P_t{\left(S\right)} k{\left(S,T\right)} + \int \dd{S'} P_t{\left(S'\right)} k{\left(S',T\right)} \varrho{\left(S\right)}. \label{eq:master}
\end{equation}

To find the equilibrium state of the model, it suffices to enforce detailed balance. Due to time-reversal symmetry, $P{\left(S|R\right)}$ is both the distribution of $S$ for particles which are about to rearrange and for particles which have just rearranged.  Thus, detailed balance requires that 

\begin{equation}
    P{\left(S|R, T\right)} = \varrho{\left(S\right)}.
\end{equation}

The equilibrium $S$ distribution at any $T$ is thus

\begin{equation}
    P_T{\left(S\right)} \propto \varrho{\left(S\right)} e^{\Delta E{\left(S\right)}/T - \Sigma{\left(S\right)}}.
    \label{eq:PTdef}
\end{equation}

Consistent with earlier results for many systems~\cite{cubuk2015identifying,sharp2018,schoenholz2017relationship,sussman2017,tah2021,tah2022fragility,ridout2022,harrington2019} , the distribution $P{\left(S\right)}$ is roughly Gaussian for both systems studied here (supplementary material).

\begin{figure}
    \centering
    \includegraphics[width=\linewidth]{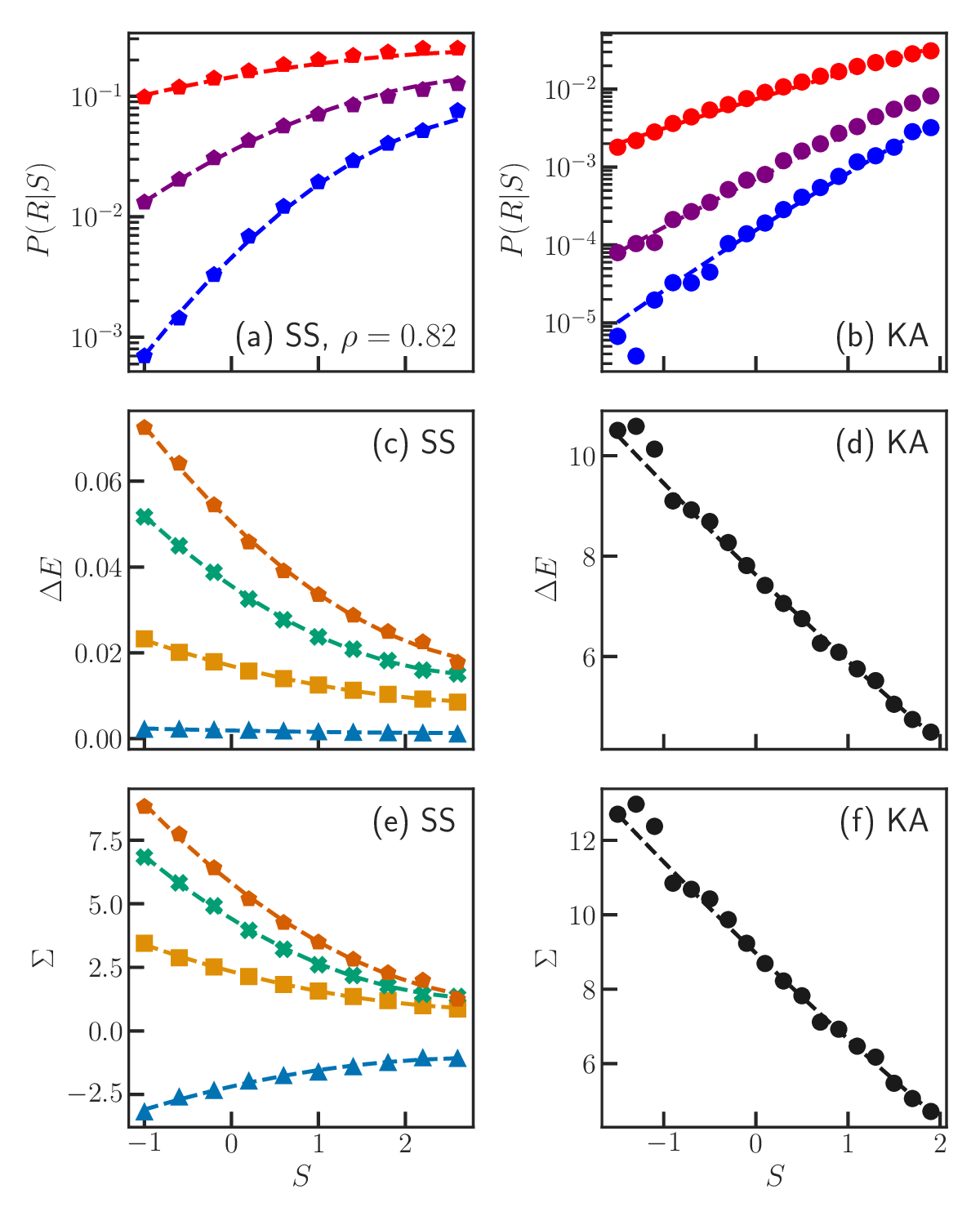}
    \caption{(a,b) $P{\left(R|S\right)}$ for both KA system ($T=0.430, 0.470,0.550$, blue to red) and SS, $\rho=0.82$ ($T=0.0045, 0.0055, 0.0065$, blue to red). (c-f)  $\Delta E{\left(S\right)}$ and $\Sigma{\left(S\right)}$, obtained by fitting $\ln{P{\left(R|S,T\right)}}$ to $\Sigma{\left(S\right)} + \Delta E{\left(S\right)}/T$, for the soft-sphere system (left column) and KALJ system (right column). In all panels, dashed lines show quadratic fits to $\Delta E{\left(S\right)}$ and $\Sigma{\left(S\right)}$ under the assumption of a (possibly negative) onset temperature (eq. \ref{eq:dE} and \ref{eq:S}), which are quite accurate in the relevant range of $S$. For SS, Orange pentagons: $\rho=0.82$, Green x's: $\rho=0.78$, Yellow squares: $\rho=0.72$, Blue triangles: $\rho=0.65$.}
    \label{fig:params}
\end{figure}

Note from Fig.~\ref{fig:params} (top) that $P{\left(R|S\right)}$ is not perfectly exponential in $S$. To ensure that $P{\left(S\right)}$ is Gaussian while accounting for some of the deviations from exponential behavior in $P{\left(R|S\right)}$, we take $\varrho{\left(S\right)}$ to be Gaussian and take

\begin{align}
    \Delta E{\left(S\right)} &= \epsilon_0 + \epsilon_1 S + \epsilon_2 S^2 \label{eq:dE} \\
    \Sigma{\left(S\right)} &= \Sigma_0 - \frac{\epsilon_1}{T_0} S - \frac{\epsilon_2}{T_0} S^2 \label{eq:S}.
\end{align}

Fig.~\ref{fig:params}(middle, bottom) shows that these functional forms for $\Delta E$ and $\Sigma$ are quite accurate.  Many systems (including our KA system and, for sufficiently large $\rho$, our SS system) exhibit an ``onset temperature'' $T_0$ where dynamical heterogeneity appears and above which $S$ is no longer predictive. The relation (eq.~\ref{eq:S}) between the coefficients for $\Delta E$ and those for $\Sigma$ is necessary for these equations, valid below $T_0$, to smoothly connect to an $S$-independent dynamics at $T_0$.  This relation even appears to hold when the inferred $T_0$ is negative~\cite{tah2022fragility}.


Eqs.~\ref{eq:PTdef}-\ref{eq:S}, with the assumption that $\varrho(S)$ is Gaussian with mean $S_0$ and variance $\sigma_0^2$, lead to the predictions

\begin{align}
\langle S \rangle &= \frac{S_0 - \gamma{\left(T\right)} \sigma_0^2}{1 - 2 \delta{\left(T\right)} \sigma_0^2}   \label{eq:mean} \\
\mathrm{Var}{\left[S\right]} &= \frac{\sigma_0^2}{1 - 2 \delta{\left(T\right)} \sigma_0^2},\label{eq:var}
\end{align}
where 
\begin{equation}
            \gamma{\left(T\right)}
 = \begin{cases} \epsilon_1 \left(\frac{1}{T_0} - \frac{1}{T} \right),  & T < T_0 \\ 0,  & T > T_0
    \end{cases} 
        \label{eq:gammadef}
\end{equation}
expresses the strength of the exponential correlation between $S$ and $P{\left(R| S \right)}$, and 

\begin{equation}
    \delta{\left(T\right)} =\begin{cases}
         \epsilon_2 \left(\frac{1}{T} - \frac{1}{T_0} \right), & T < T_0 \\ 0, & T > T_0
    \end{cases}
    \label{eq:deltadef}
\end{equation}

giving a quadratic correction to $\ln{P{\left(R|S\right)}}$.

Note that above $T_0$, $\langle S \rangle$ is predicted to no longer depend on $T$. Recall that $T_0$ also marks the temperature above which $E_{\mathrm{IS}}$ does not depend on $T$~\cite{sastry1998}. Since $E_{\mathrm{IS}}$ and $S$ both reflect the same IS structure, this model thus links the thermodynamic and dynamical definitions of $T_0$.

Now consider the overlap function and dynamical heterogeneity within the model.  The mean overlap is 

\begin{equation}
Q{\left(t\right)} =\frac{1}{M} \sum_i q_i{\left(t\right)},
\end{equation}

where $q_i{\left(t\right)}=1$ if site $i$ has not rearranged before time $t$ and $q_i{\left(t\right)}=0$ if it has. Correlations are measured by 

\begin{equation}
    \chi_4{\left(t\right)} \equiv N \, \mathrm{Var}{\,Q{\left(t\right)}}.
\end{equation}

Because the sites are treated as independent,  

\begin{equation}
\overline{Q}{\left(t\right)} = \int \dd{S} P{\left(S\right)} e^{-k{\left(S\right) t}}. \label{eq:q}
\end{equation}

This lacks a closed-form expression, even if we set $\delta=0$ (eq.~\ref{eq:deltadef}).
Because sites are independent, there are no dynamical correlations (beyond grouping together $m$ particles into one site) and $\chi_4{\left(t\right)}$ is trivial. Using only this independence, with no assumptions about $P{\left(S\right)}$ or the transition rates, one may easily show that\cite{nishikawa2022}

\begin{equation}
    \chi_4{\left(t\right)} \equiv N \left(\overline{Q^2} - \overline{Q}^2 \right)= m \left(\overline{Q{\left(t\right)}}- \overline{Q{\left(t\right)}}^2\right). \label{eq:chi}
\end{equation}

The quantity $\chi_4$ peaks at some time $\tau_\chi$ at a value of $\chi_4^*$; it follows from Equation \ref{eq:chi} that

\begin{align}
    \overline{Q}{\left(\tau_\chi\right)} &= \frac{1}{2} \label{eq:tauchi}\\
    \chi_4^* &= \frac{m}{4}. \label{eq:peak}
\end{align}

Equation \ref{eq:peak} is in agreement with previous counting arguments stating that $\chi_4^*$ should be proportional to the number of particles $m$ in a ``mobile domain''.~\cite{abate2007topological}

Clearly, this model cannot account for the growth of $\chi_4^*$ with $1/T$. Doing so requires a model with spatial correlations in $S$, which will be discussed in a future manuscript.



The first step is to use the trap-like model to calculate the relaxation time as a function of $T$. We calculate two different relaxation times. The first, $\tau_Q$, corresponds to the time at which $Q(t)$ drops to $1/e$. The second, $\tau_\chi$, is the position of the peak of $\chi_4(t)$. These timescales are generally comparable\cite{abate2007topological,lacevic2003,tah2021understanding}. Analysis of eq. \ref{eq:q} shows

\begin{align}
    \tau_Q k{\left(\langle S \rangle, T\right)} &= g_Q{\left(\gamma{\left(T\right)} \sigma_S, \delta{\left(T\right)} \sigma_S^2, \delta{\left(T\right)} \sigma_S \langle S\rangle  \right)} \label{eq:qscale}\\
    \tau_\chi k{\left(\langle S \rangle, T\right)}  &=  g_\chi{\left(\gamma{\left(T\right)} \sigma_S, \delta{\left(T\right)} \sigma_S^2, \delta{\left(T\right)} \sigma_S \langle S\rangle  \right)}. \label{eq:chiscale}
\end{align}

Thus, the basic timescale of relaxation remains $1/k{\left(\langle S \rangle, T\right)}$ (see eq.~\ref{eq:rate}), in agreement with previous observations that $\tau \approx 1/P{\left(R|\langle S \rangle\right)}$~\cite{schoenholz2017relationship,landes2020attractive,tah2021}, but even in the trap-like model the variance of $S$ gives corrections to this relation. It is therefore important to calculate $g_\chi$ and $g_Q$ in eq.~\ref{eq:chiscale}, which we do by evaluating eq.~\ref{eq:q} numerically.


In predictions discussed below, we use the specific values of $\gamma{\left(T\right)}, \delta{\left(T\right)}$ predicted for each system. We find that $g_Q$ is an increasing function of $\gamma$ and $\delta$, while $g_{\chi}$ depends very little on $T$. Since $\gamma$ and $\delta$ increase with cooling, this theory thus predicts that $\tau_Q$ exhibits a slightly stronger super-Arrhenius growth than $\tau_\chi$. To see why, consider a system where half the sites are very fast and half are very slow.  $\chi_4$ peaks when half of the regions rearrange; therefore it peaks near the fast timescale.  On the other hand, $\tau_Q$, being defined by $Q{\left(\tau_Q\right)} = 1/e \approx 0.37$, will not be reached until many slow regions have also relaxed.



\section{Relaxation times and fragility}

We now test the trap-like model against simulation data for both the KA and SS systems.  For each system, we define the underlying softness distribution, $\rho{\left(S\right)}$, to be Gaussian with the mean and variance of $S$ obtained from simulations at a single high temperature that is close to $T_0$ when it is positive.

\begin{figure}
    \centering
    \includegraphics[width=\linewidth]{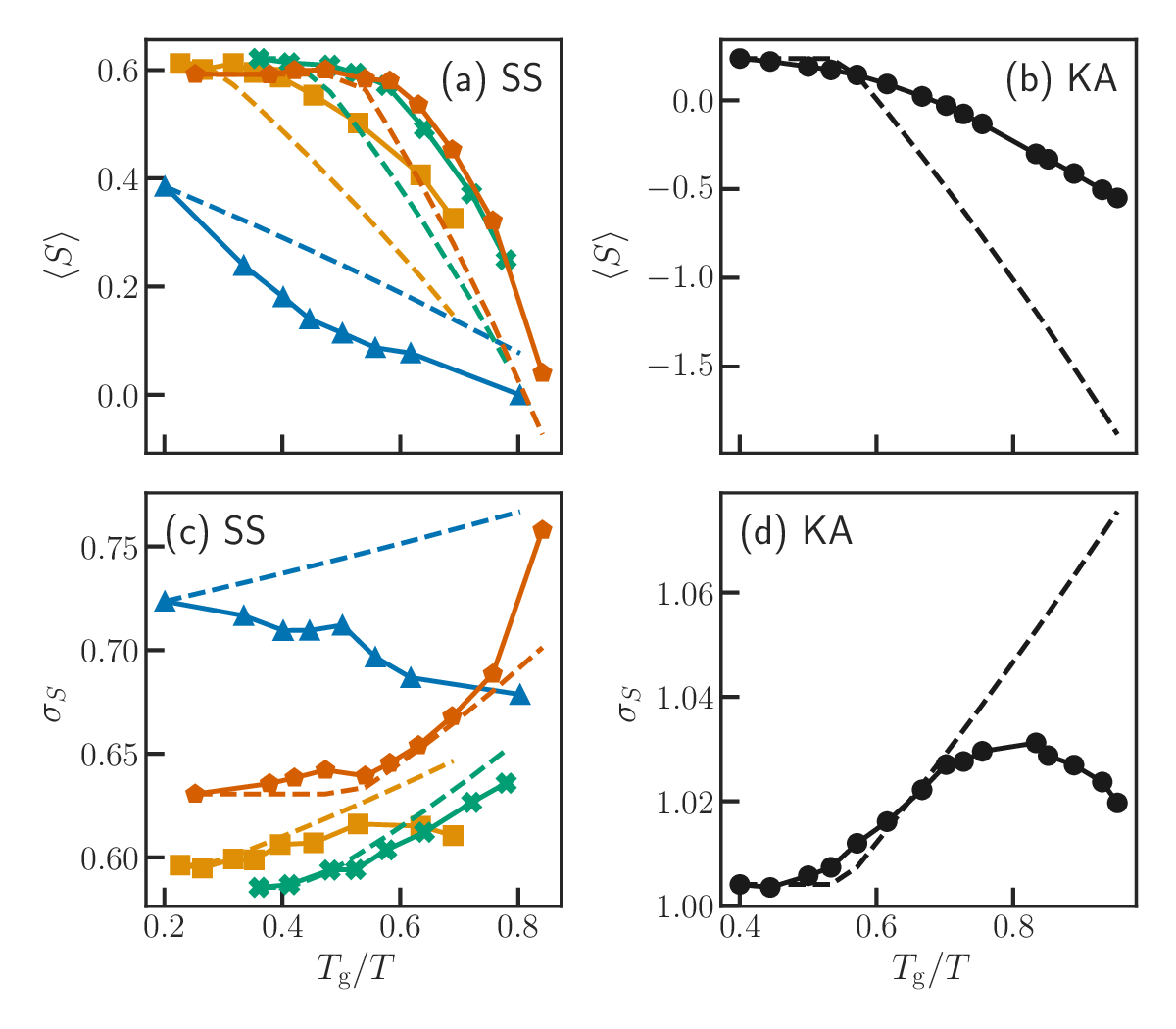}
    \caption{Mean softness $\langle S \rangle$, and standard deviation $\sigma_S$, vs. inverse temperature.  Dashed lines show the prediction of the trap-like model. (a): $\langle S\rangle$ in the SS system. The model overestimates the change in $\langle S\rangle$ at most densities, while underestimating it at $\rho=0.65$. (b): $\langle S\rangle$ in the KALJ system. In this system, the dependence of $\langle S \rangle$ on $\beta$ is strongly overestimated. (c,d): $\sigma_S$ in (SS, KA) systems, showing very weak $T$-dependence. For SS, Orange: $\rho=0.82$, Green: $\rho=0.78$, Yellow: $\rho=0.72$, Blue: $\rho=0.65$.}
    \label{fig:SvsBeta}
\end{figure}

\begin{figure}
    \centering
    \includegraphics[width=\linewidth]{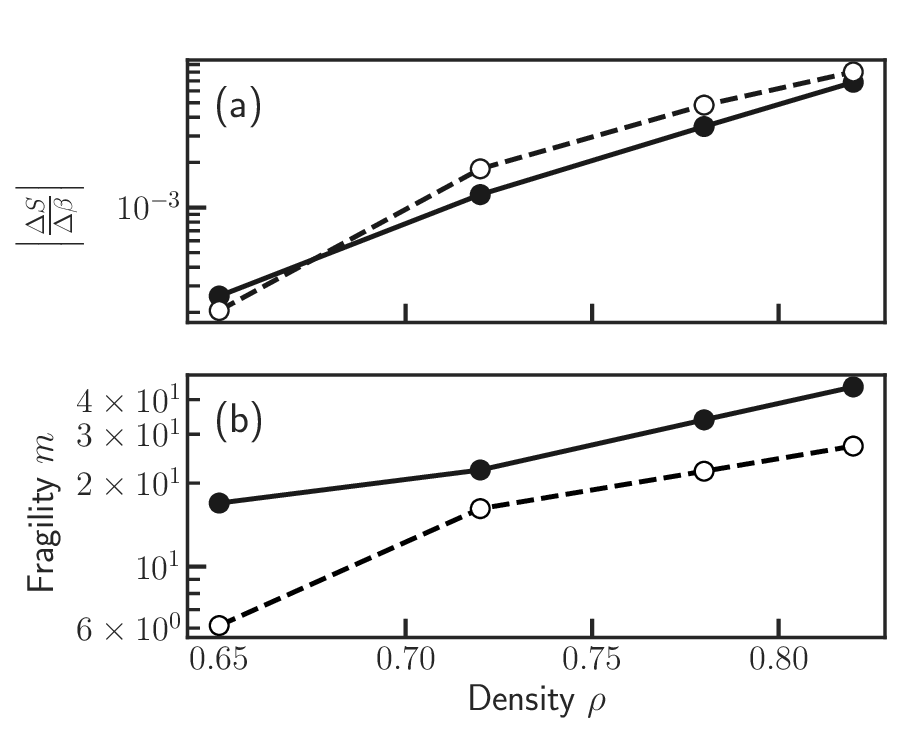}
    \caption{(a) Change of mean softness between highest and lowest temperatures, per unit inverse temperature, as a function of density $\rho$. Solid: MD data, dashed: trap-like model.  Both the trend with density and the magnitude of the change are predicted surprisingly well by the model.  (b) Fragility index $m$ as a function of $\rho$. The trap-like model underestimates the fragility at all densities, but correctly predicts that fragility increases with density.}
    \label{fig:fragility}
\end{figure}

As shown in eq.~\ref{eq:mean}, the model predicts that $\langle S \rangle$ decreases with $1/T$ more quickly for larger $\epsilon_1$.  In the SS system, Fig.~\ref{fig:params} shows that $\epsilon_1$ increases with density $\rho$; accordingly, fig.~\ref{fig:SvsBeta}(a) shows a stronger dependence of $\langle S \rangle$ on $T_g/T$ at larger $\rho$ (corresponding to increasing fragility~\cite{tah2022fragility}); this correct trend with density is quantified in fig.~\ref{fig:fragility}(a). The agreement is imperfect, and the curvature of $\langle S\rangle$ vs. $1/T$ is not predicted well.  In the KA system (fig. \ref{fig:SvsBeta}(b)), the change of $\langle S \rangle$ with $T$ is overestimated.

In the trap-like model, the standard deviation of $S$ depends weakly on temperature.  Fig. \ref{fig:SvsBeta}(c,d)) shows that this is true in the MD data, and the prediction is reasonably accurate at moderate $T$, except at $\rho=0.65$ (SS) .

We now turn to the predictions of relaxation times. Our model only considers transitions of the system between inherent states, neglecting the thermal fluctuations within a cage. Thus, the most direct comparison may be made to the $\overline{Q}$ or $\chi_4$ computed on IS positions.  We find that the IS $\chi_4$ in the high-density soft-sphere system behaves poorly at high $T$ (approaching $T_0$), and thus here choose to focus on $\tau_Q$ computed from IS positions. $\tau_\chi$ at $\rho=0.65$ is discussed in the supplement.

\begin{figure}
    \centering
    \includegraphics[width=\linewidth]{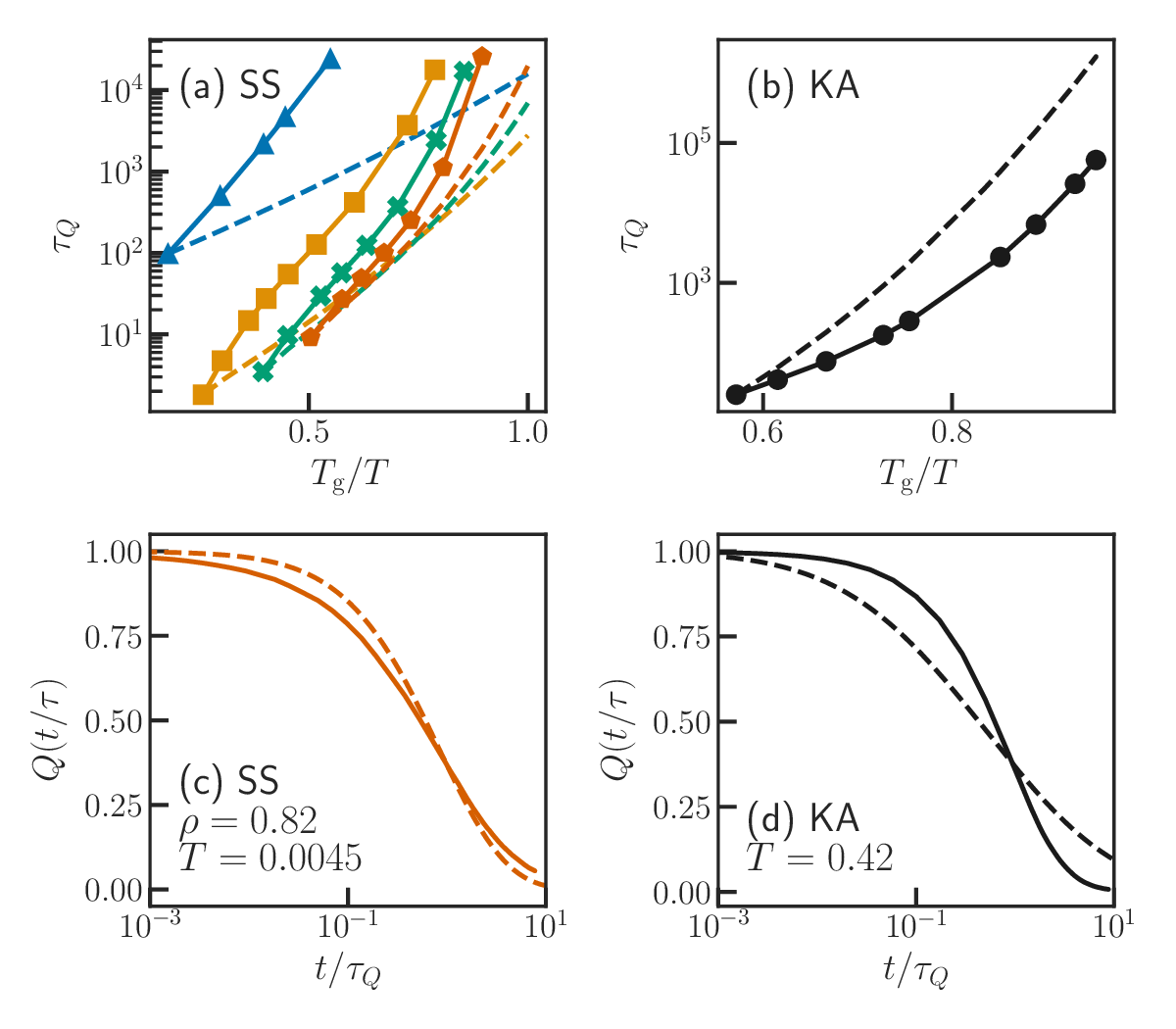}
    \caption{(a,b) Prediction of $\tau_Q$  from the trap-like model for all systems. (b) As discussed in the text, $\tau$ is overestimated in the KA system , while (a) in the SS system $\tau$ and the fragility $m$ are underestimated at all densities, although the trend with density is correct (fig. \ref{fig:fragility} (b.)).  (c) The model underestimates the stretching of $Q{\left(t\right)}$ for the SS system, while (d) overestimating it for KA. Colors and symbols as in fig. 1, dashed lines are trap predictions.}
    \label{fig:tau}
\end{figure}

The predicted $\tau_Q$ is shown for both systems in fig. ~\ref{fig:tau}(a,b). It is clear that not only does the trap-like model underestimate the relaxation time $\tau_Q$ for the SS system and overestimate it for the KA system, it also strongly underestimates fragility for SS. To quantify this, we follow Angell~\cite{bohmer1992correlations} to define the fragility index $m$ as



\begin{equation}
m = \left.\dv{\ln{\tau}}{T_\mathrm{g}/T}\right|_{T_\mathrm{g}}.
\end{equation}

In the soft-sphere system, we take $T_{\mathrm{g}}$ to be the temperature at which a parabolic-law extrapolation~\cite{elmatad2009} of $\tau$ reaches $10^7$ in the MD data; for the simulations, $m$ is computed from this extrapolation.

In Fig.~\ref{fig:fragility}(b), we compare the fragility $m$ computed from the MD data to the trap prediction for the soft-sphere system. Interestingly, the trend with density is very well captured by the model even though $m$ is consistently underestimated. 
Past results have emphasized the apparent correlation between $\tau$ and $1/P{\left(R|\rangle S \rangle\right)}$ ~\cite{schoenholz2016structural,schoenholz2017relationship,landes2020attractive}:
\begin{equation}
    \tau \sim 1/P{\left(R|\langle S{\left(T\right)} \rangle, T\right)}. 
    \label{eq:tauprrelation}
\end{equation}
 We have recently noted, however, that careful analysis shows some deviation from this relation at low $T$ in the soft-sphere system \cite{tah2022fragility}.  Furthermore, we have seen above that even in the traplike model, with no facilitation or other correlations between particles, this simple relation is \textit{not} expected to hold for $\tau_Q$.  The consistent underestimation of $m$, however, indicates that even with the traplike model's correction to eq.~\ref{eq:tauprrelation}, a model based on independent relaxation of particles with rates derived from $P{\left(R|S,T\right)}$ still underestimates fragility for the SS system.

In the KA system, taking $T_\mathrm{g}=0.40$, we find $m$ is slightly overpredicted by the model ($m=45.2$ vs. $m=37.2$).

The shape of $Q{\left(t\right)}$ is not well predicted by the trap-like model. Fig. \ref{fig:tau}(c,d) show that, at low $T$, the model underestimates the stretching of $Q$ in the SS system while overestimating it in the KA system (details in supplement).



\section{Aging}

\begin{figure}
    \centering
    \includegraphics[width=\linewidth]{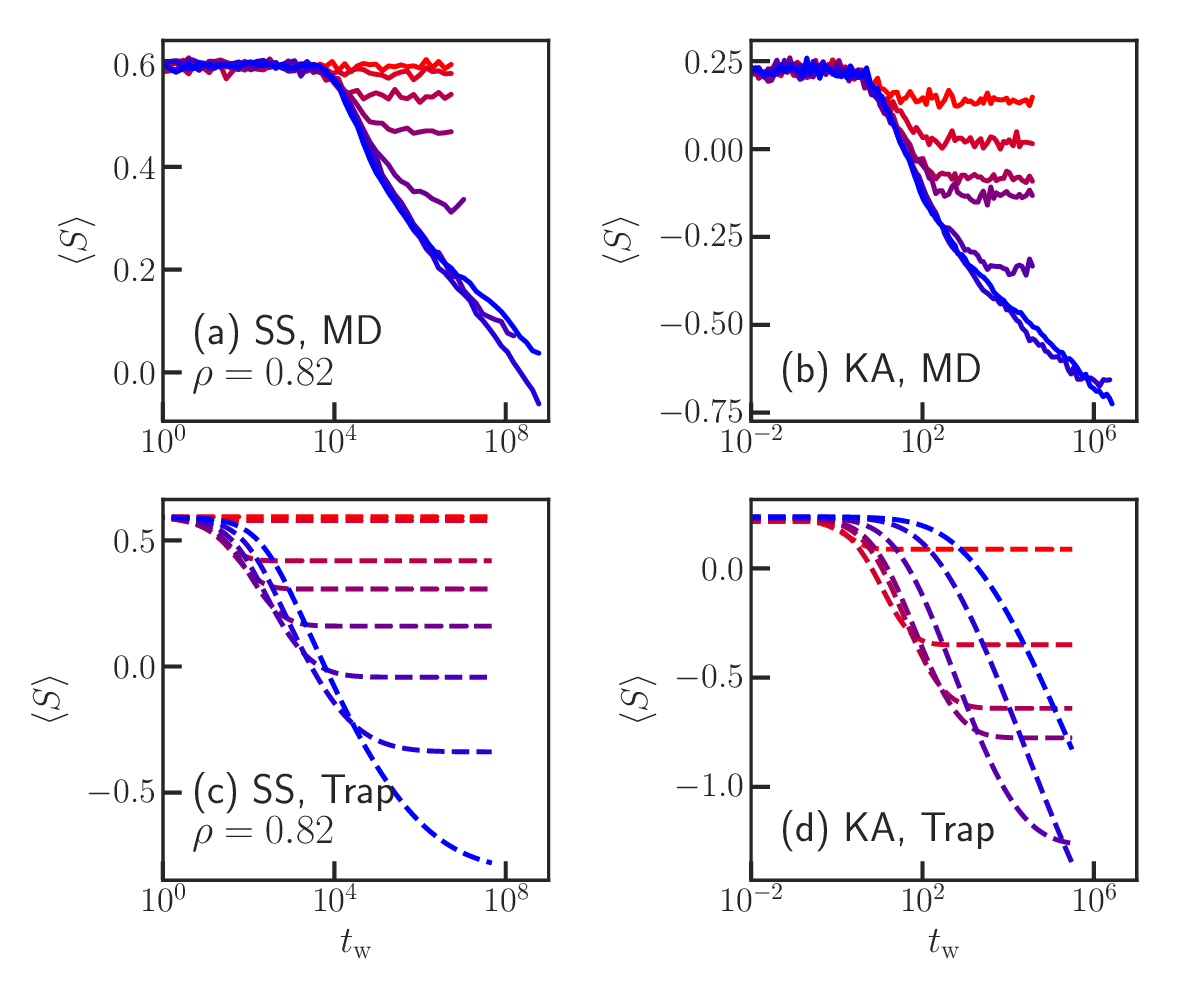}
    
    \caption{Evolution of $\langle S \rangle$ during aging from $\hot$ to $\cold$.   (a): MD data,  SS at $\rho=0.82$; $\hot=\maxcoldSS$, $\cold=0.008$, $0.007$, $0.006$, $0.0055$, $0.005$,$0.0045$, $0.004$, $0.0035$ (red to blue). (b):  MD data, KA system. $\hot=1.00$, $\cold=0.70$, $0.60$, $0.55$, $0.53$, $0.47$, $0.40$, $0.35$ (red to blue).  Both systems have an intermediate time regime where $\langle S \rangle$ ages logarithmically and the curves for different values of $\cold$ roughly overlap.  (c): Trap-like model, SS system at $\rho=0.82$. (d): Trap-like model, KA system.  As discussed in main text, the model fails to capture the collapse of the curves. }
    \label{fig:aging}
\end{figure}

In our MD simulations, we study aging by preparing a system at a high temperature $\hot$, then connecting it to a cold thermostat at $\cold$ and watching its structure and relaxation time evolve with age or waiting time since the quench, $t_w$~\cite{kob1997aging,schoenholz2017relationship}. In principle, understanding an aging system requires knowing its entire history. However, earlier work~\cite{schoenholz2017relationship} found that $S$ behaves as an ``internal variable" that captures the relevant history dependence so that the system can be characterized by the usual state variables with the addition of $S$~\cite{schoenholz2017relationship}. In particular, the relation between structure and dynamics, as encapsulated by $P{\left(R|S,T\right)}$, was found to be independent of age.

$\langle{\left(t_w\right)}\rangle$ has another remarkable behavior. The top row of Fig.~\ref{fig:aging} shows MD data for (a) SS, $\rho=0.82$ and (b) KA, for systems prepared at fixed $\hot$ and quenched to different $\cold$. The KA data agree with previously published results ~\cite{schoenholz2017relationship}. In both systems, the $\langle S{\left(t_w\right)} \rangle$ curves for \textit{different} $\cold$ lie on top of each other in the logarithmic aging regime.

To see why this is surprising, consider two systems at the same value of $\langle S \rangle$, but connected to thermostats with different values of $\cold$. Particles of a given $S$ rearrange less in the colder system\cite{schoenholz2017relationship}. Thus, it seems that $\langle S \rangle$ would evolve more slowly with age for the colder system.  The collapse of $\langle S \rangle$ for different $\cold$ shows that this is not the case---somehow, although there are fewer rearrangements at lower $\cold$, they are more effective in lowering $\langle S \rangle$ so that $\langle S \rangle$ evolves at the same rate for different $\cold$.

We study aging in our trap-like model numerically by solving the master equation \ref{eq:master} for a fine discretization of $S$ and $t$. We find that $\langle S \rangle$ does have a logarithmic aging regime (bottom row, Fig.~\ref{fig:aging}) for both systems, as seen in MD (top row, Fig.~\ref{fig:aging}). However, fig.~\ref{fig:aging}(c,d) indeed shows that in the trap-like model, $\langle S \rangle$ decreases more slowly for colder systems. Thus, the collapse seen in MD data (Fig.~\ref{fig:aging}(a,b)) is not predicted by the trap-like model, showing that the collapse of $\langle S{\left(t_w\right)} \rangle $ must arise from correlations neglected in this model.

\section{Discussion}

The softness formulation of the glass problem lends itself naturally to a traplike model since there is a well-defined energy barrier to rearrangements that depends on softness, $\Delta E(S)$. The traplike model neglects the effects on energy barriers of correlations between particles, so the degree to which the traplike model captures, or does not capture, observed behavior gives us insight into the importance of correlations and facilitation. 

The traplike model is remarkably successful in capturing the rate of change of mean softness, $\abs{\frac{\Delta S}{\Delta \beta}}$, with inverse temperature $\beta$, as a function of density $\rho$ in soft-sphere systems (fig.~\ref{fig:fragility}(a.)). Likewise, it captures quite well the trend of increasing fragility $m$ with $\rho$ in those systems. 

However, the traplike model tends to underestimate $\langle S \rangle$, predicting a stronger decrease with decreasing $T$ than is observed (Fig.~\ref{fig:SvsBeta}, top row), except for SS, $\rho=0.65$.


We turn now to predictions for the relaxation time.  The traplike model underpredicts/overpredicts the relaxation time $\tau_Q$ for the SS/KA systems, respectively (fig.~\ref{fig:tau}). It is easy to see why the traplike model might predict a relaxation time that is too high. First, as discussed above, $\left| d \langle S \rangle / d T \right|$ is overestimated in the KA system.  Thus, the predicted energy barriers are too high at low $T$, giving prediction of $\tau$ that is too high. Second, in reality particles' $S$ changes when their neighbours rearrange. This is a form of facilitation, which should shorten the relaxation time~\cite{schoenholz2016structural}. Consistent with this picture is the excessive stretching of $Q{\left(t\right)}$ predicted by the trap-like model for the KA system (fig. \ref{fig:tau}(d)), which should be reduced by facilitation allowing the hardest particles to relax sooner. 

The quality of $S$ as a predictor has an important effect on the predictions of the trap-like model. A better predictor should yield a broader range of energy barriers, and thus, in the trap-like model, predict greater heterogeneity and fragility. The underprediction of fragility $m$ in the SS model could result from this effect.  Note that $S$ is a less accurate predictor of rearrangements in the SS model than in the KA model~\cite{tah2022fragility}. This is consistent with the fact that the predicted $m$ is too low for the SS system but almost correct for the KA system, as well as the insufficient stretching of $Q{\left(t\right)}$ predicted for the SS system (fig.~\ref{fig:tau}(c)).


We have only compared the trap-like model to simulations that exhibit behavior ranging from Arrhenius to super-Arrhenius. The trap-like model predicts a decrease in $\langle S \rangle $ with $T$\footnote{Unless there exists a system in which $T_0 > 0$ but it is the entropic effect of structure, rather than the energetic effect, which dominates.}, which would appear to inevitably lead to super-Arrhenius growth of the relaxation time. However, there are systems (such as the SS system at lower densities) that exhibit sub-Arrhenius growth~\cite{berthier2009compressing}.  We note that sub-Arrhenius behavior could originate in the trap-like model from the prefactor (an attempt frequency) in eq.~\ref{eq:rate}, which must depend on the mean-squared velocity, which scales as $\sqrt{T}$. This prefactor can lead to sub-Arrhenius growth if $\Delta E(S)/T$ is small and thus the contribution of changes in $\langle S \rangle$ to $\tau$ is very small.  One should check for this prefactor in other models where $P{\left(R|S,T\right)}$ is known to correlate with a sub-Arrhenius relaxation time~\cite{tah2021}.  Similarly, the trap-like model cannot account for a fragile-to-strong crossover with cooling (as suggested for the KA system in \cite{das2022}) without e.g. a non-Gaussian distribution of traps. Note that in an MD model of silica with a fragile-to-strong crossover, $P{\left(S\right)}$ is clearly non-Gaussian \cite{cubuk2020}.

Recent work indicates that structure retains small predictiveness of dynamics above the nominal onset temperature \cite{obadiya2022fluid}, suggesting small corrections to eqs. \ref{eq:gammadef},\ref{eq:deltadef} near $T_0$. This will smooth the crossover in $\langle S \rangle$ predicted by the trap-like model. This is consistent with the fact that $T_0$ marks a crossover in IS energy, not a sharp transition \cite{sastry1998}.



\section{Conclusions}

Past work has suggested that a machine-learned structural variable, softness $S$, can lead to a significant simplification of our understanding of the dynamics of supercooled liquids  by introducing a distribution of local energy barriers to rearrangements.  In this picture, fragility is connected to the temperature dependence of $\langle S \rangle$. We have shown how a simple model of the dynamics of $S$, formally similar to the trap model, qualitatively explains the dependence of $\langle S \rangle$ on $T$, predicting the relative fragility of different glassformers within the same family.   We have also shown how this simple model severely fails to describe interesting features of the dynamics, including aging of $\langle S \rangle$, suggesting a clear need for a model which includes facilitation through the effects of rearrangements on the softness of neighbouring particles.

\section{Acknowledgements}
We thank Horst-Holger Boltz for insightful comments that inspired this line of work, and Rahul Chacko, Tomilola M. Obadiya, and Daniel M. Sussman for helpful discussions. This work was supported by the Simons Foundation via the ``Cracking the glass problem'' collaboration (\#454945, SAR, IT and AJL), the Simons Investigator program (\#327939, IT and AJL) and the hospitality of the Flatiron Institute (AJL).

\bibliography{bibliography}

\pagebreak
\setcounter{equation}{0}
\setcounter{figure}{0}
\setcounter{table}{0}
\setcounter{page}{1}
\makeatletter
\renewcommand{\theequation}{S\arabic{equation}}
\renewcommand{\thefigure}{S\arabic{figure}}


\section{Training of Softness}

Our training of softness roughly follows references \cite{schoenholz2016structural, tah2022fragility}.  

In both systems we use the radial structure functions 

\begin{equation}
G{\left(i;  \mu, \sigma\right)} = \sum_{ j | r_{ij} < \sigma_{\mathrm{max}}} e^{-\left(r_{ij} - \mu\right))^2 /\sigma^2 }.
\end{equation}

In the KALJ system we take $\sigma = 0.05 \sigma_{AB}$ and $\mu = 0.05 \sigma_{AB}, 0.1 \sigma_{AB}, \dots, 5 \sigma_{AB}$. In the soft-sphere system we take $\sigma = 0.1 \sigma_{AA}$ and $\mu = 0.1 \sigma_{AA}, 0.2 \sigma_{AA}, \dots, 5 \sigma_{AA}$.  In both cases $\sigma_{\mathrm{max}}$ is equal to the largest value of $\mu$.

We also use the angular structure functions

\begin{align}
&\Psi{\left(r; \xi, \lambda, \zeta \right)} =\nonumber  \\ &\sum_{ j,k | r_{ij}, r_{jk} < \sigma_{\mathrm{max}}} e^{-\left(r_{ij}^2 + r_{jk}^2 + r_{ik}^2\right)/\xi^2 } \left( 1 + \lambda  \cos{\theta_{ijk}}\right)^{\zeta},
\end{align}

with the parameters $\xi$, $\lambda$, and $\zeta$ as in \cite{cubuk2015identifying}.  Thus, the only difference from \cite{cubuk2015identifying,schoenholz2016structural} is that in the KA system we use more finely-spaced radial structure functions.

For the KALJ system, a training set is constructed  at $T=0.470$. Particles are labelled as rearranging if they have $p_{\mathrm{hop}} > 0.6$, and labelled as non-rearranging if they have $p_{\mathrm{hop}} < 0.01$ for a time of $1000 \tau \approx 2 \tau_{\alpha}$. Equal-sized random samples are taken of rearranging and non-rearranging particles. The definition of $p_{\mathrm{hop}}$, and the use of IS positions to compute both both $p_{\mathrm{hop}}$ and the structure functions, are the same as in \cite{schoenholz2016structural}.

For the soft-sphere system, a training set is constructed using an equal mixture of data from $\rho=0.65, T=0.00065$ and $\rho=0.82, T=0.0045$.  These state points have similar relaxation times, making it reasonable to construct a mixed training set which draws rearranging and non-rearranging examples from them equally.  Particles are labelled as rearranging if they have $D^2_{\mathrm{min}} > D^2_{\mathrm{min}, 0}$ over a window of $\Delta t = 12$, and labelled as non-rearranging if they have $D^2_{\mathrm{min}} < D^2_{\mathrm{min}, 0}$  over a time window of $10 \tau_{\alpha}$. For each density $D^2_{\mathrm{min}, 0}$, is chosen such that $1\%$ of particles are rearranging. The structure functions are computed using inherent structure positions. Aside from defining structure functions using IS positions, this matches \cite{tah2022fragility}.

In both cases the SVM is trained using a squared-hinge loss with a penalty parameter $C=1$, which results in a cross-validation accuracy of 93\% in the KALJ system and 83\% in the soft-sphere system. As in \cite{tah2022fragility}, the soft-sphere hyperplane achieves similar test accuracy on intermediate densities which were not included in the training set, at temperatures chosen to have a similar relaxation time.  

\section{$P{\left(S\right)}$ for all systems}

\begin{figure}
\includegraphics[width=\columnwidth]{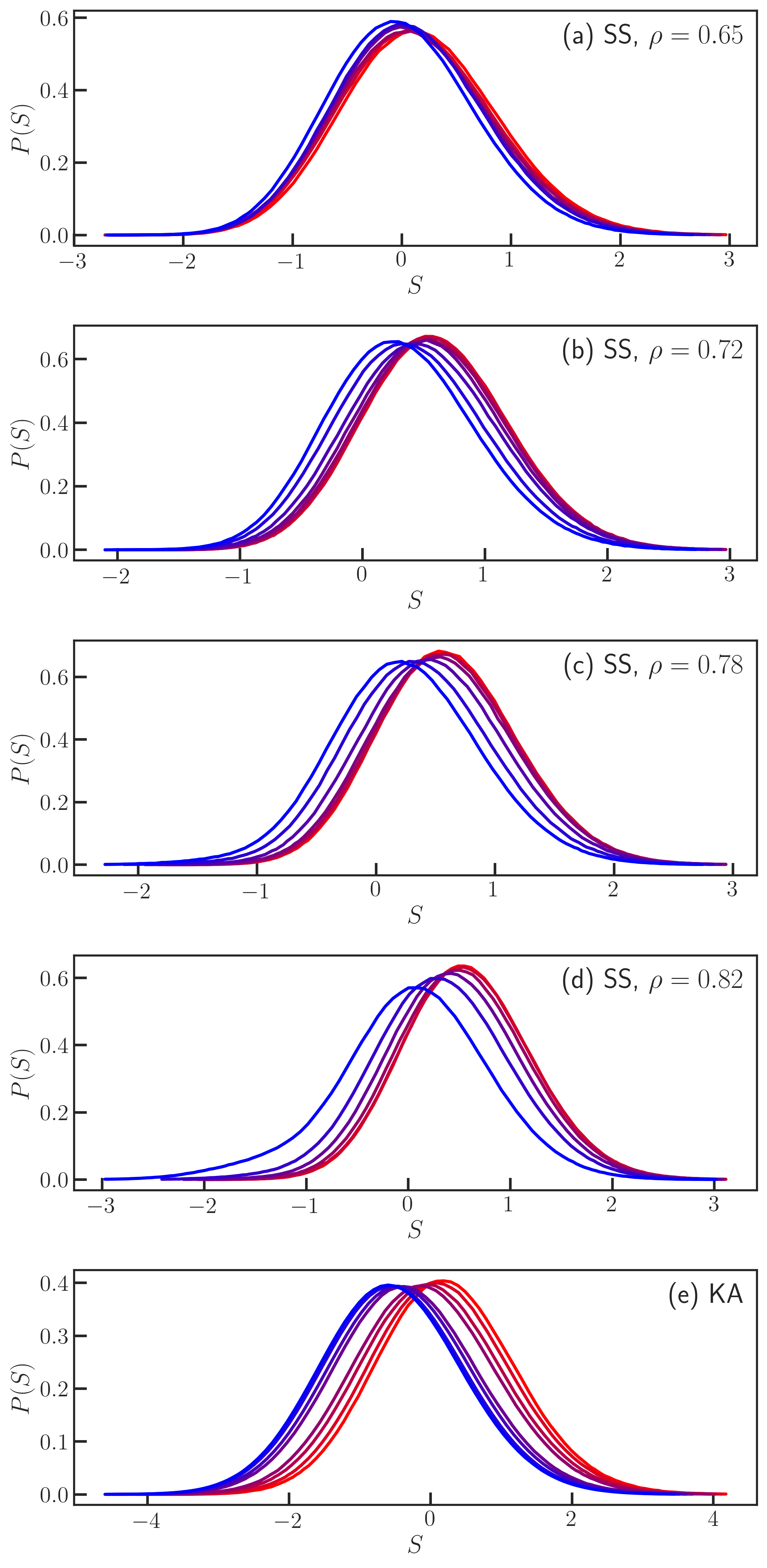}
\caption{Distribution of $P{\left(S\right)}$ for all systems is approximately Gaussian. From high temperatures near $T_0$ (red) to low temperatures (blue). Temperatures: SS $\rho=0.65$: $T=0.001$,$0.0009$,$0.0008$,$0.00072$,$0.00065$,$0.00050$, $\rho=0.72$: $T=0.006$,$0.005$,$0.0045$,$0.004$, $0.0035$, $0.003$, $0.0025$, $0.0023$ , $\rho=0.78$: $T=0.007$, $0.006$, $0.0055$, $0.005$, $0.0045$, $0.004$, $0.0037$, $\rho=0.82$: $T=0.007$, $0.0065$, $0.006$, $0.0055$, $0.005$, $0.0045$, KA: $T=1.0$, $0.70$, $0.60$, $0.55$, $0.47$, $0.45$,  $0.43$, $0.42$}\label{fig:supp_ps}
\end{figure}

Figure \ref{fig:supp_ps} shows $P{\left(S\right)}$ for all systems, which is roughly Gaussian in agreement with past work.

\section{$P(R|S)$ for all systems}

As in \cite{schoenholz2016structural,tah2022fragility}, rearranging particles were defined either by having $p_{\mathrm{hop}} > 0.2$ (KA) or $D^2_{\mathrm{min}} > D^2_{\mathrm{min}, 0}$ (SS).

\begin{figure}
\includegraphics[width=\columnwidth]{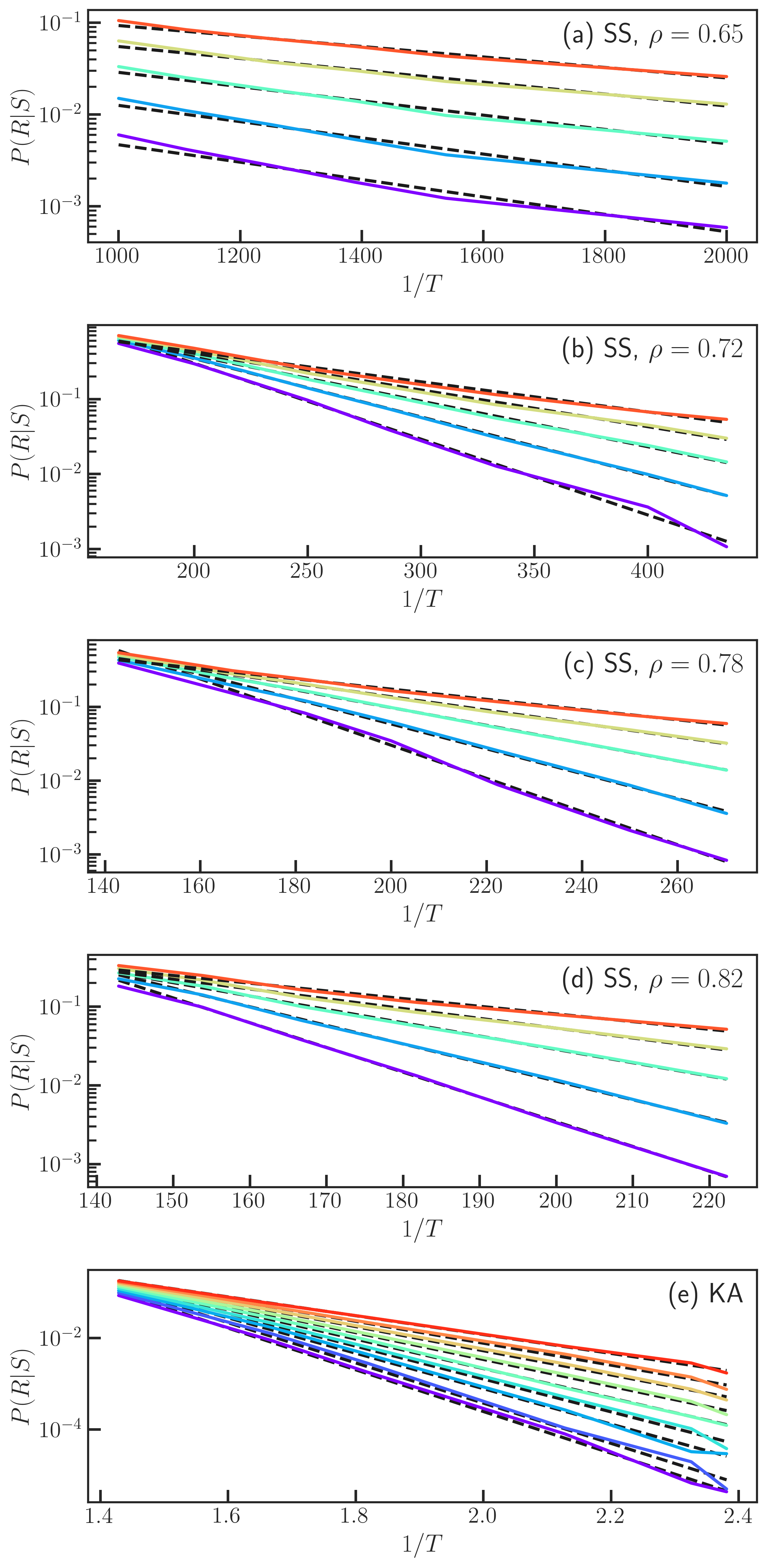}
\caption{Arrhenius fits to $P{\left(R|S\right)}$ for $T<T_0$. Each curve is a different softness bin. (SS: $S=-1$, $-0.2$, $0.6$, $1.4$, $2.2$. KA: $S=-1.5$, $-1.1$, $\cdots$, $1.7$)}\label{fig:supp_arr}
\end{figure}

\begin{figure}
\includegraphics[width=\columnwidth]{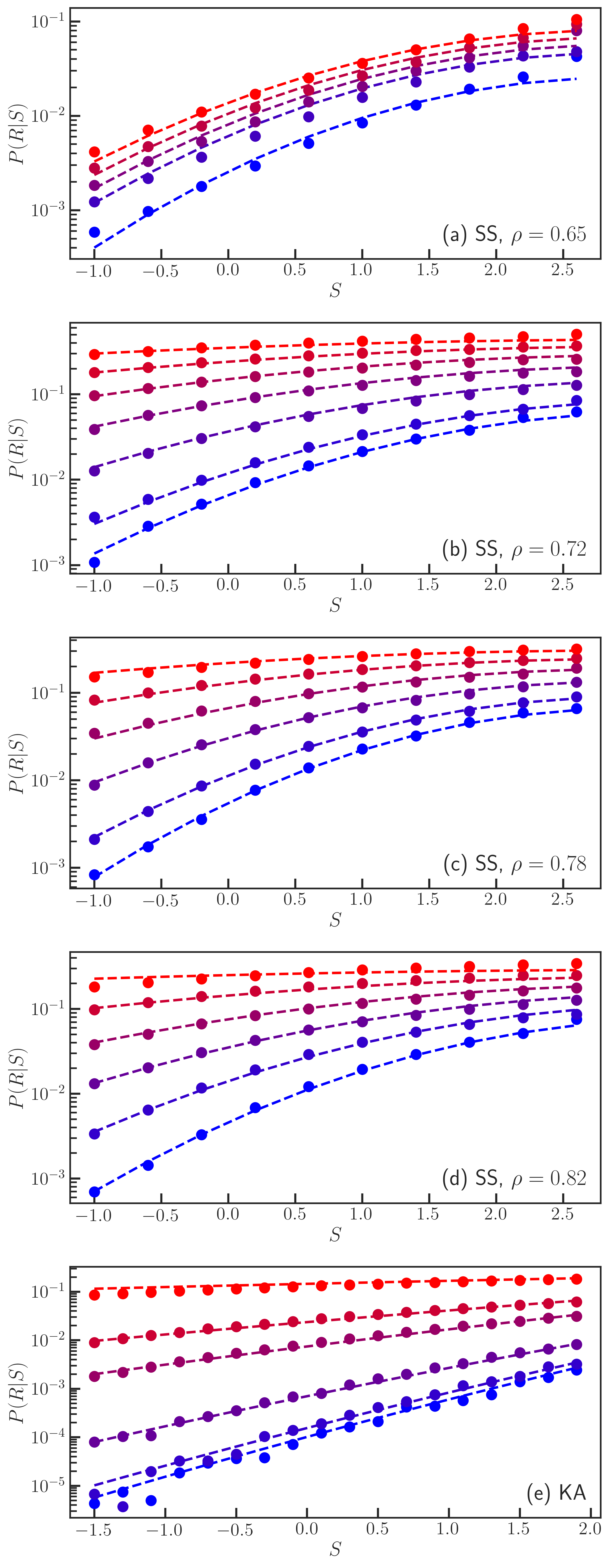}
\caption{ $P{\left(R|S\right)}$ vs. $S$, compared to the final Arrhenius fit with $\Delta E(S)$, $\Sigma(S)$ quadratic functions of $S$ (main text eq. 8, 9), constrained to become independent of $S$ at a temperature $T_0$.  Temperatures as in fig. \ref{fig:supp_arr} }\label{fig:supp_prs}
\end{figure}

Figure \ref{fig:supp_arr} shows Arrhenius fits to $P{\left(R|S,T\right)}$, used to determine the $\Delta E{\left(S\right)}$, $\Sigma{\left(S\right)}$ shown in main-text figure 1. Note the deviation at high-$T$; these high-$T$ points are not included in the fit. Figure \ref{fig:supp_prs} compares the final fit (forcing $\Delta E$ and $\Sigma$ to be quadratic in $S$ and connected via an onset temperature $T_0$) to $P{\left(R|S\right)}$ at all $T < T_0$, showing reasonable agreement.

\section{Rescaling of rearrangement rate}

Predictions of how $P(S)$ depends on $T$ are independent of any rescaling of $k(S,T)$ by a constant factor, e.g. a timescale.  However, comparison of relaxation times to simulation requires fixing an absolute timescale.

Converting $P(R|S)$ into a rearrangement rate requires dividing by a timescale. In the case of rearrangements defined using $D^2_{\mathrm{min}}$ this is simply the time interval over which $D^2_{\mathrm{min}}$ is computed. In the case of $p_{\mathrm{hop}}$, however, it must be the average duration of time for which a single rearrangement results in $p_{\mathrm{hop}}$ being above the threshold. Note that this is different than the conversion timescale used in past work \cite{schoenholz2016structural}.

Secondly, the overall rate of rearrangements may be different when measured using $D^2_{\mathrm{min}}$ or $p_{\mathrm{hop}}$ than using displacement, as they tend to filter out many events. 

To account for both of these effects, we rescale $P(R)$ at all temperatures by a temperature-independent factor decided by matching $\tau$ at a single high temperature.

Finally, we note that, if $P{\left(R |S, T\right)} \approx 1$ it would seem more correct to take e.g. $P{\left(R |S, T\right)} = 1 -e^{-k{\left(S,T\right)} \tau_R}$; we find, however, that if $P{\left(R |S, T\right)}$ is close to $1$, which only occurs close to the onset temperature $T_0$, then $P{\left(R\right)}$ deviates from Arrhenius behavior even with this correction; we do not include such high temperatures in the fit.

\section{$\tau_Q$, $\tau_\chi$, and the shape of $Q{\left(t\right)}$}

As discussed in the main text, the trap-like model (1) predicts that $\tau_\chi = \tau_{1/2}$, defined as $Q{\left(\tau_{1/2}\right)} = 1/2$, and (2) makes a prediction for the ratio $\tau_Q / \tau_{1/2}$ which grows as temperature decreases.

\begin{figure}
\includegraphics[width=\columnwidth]{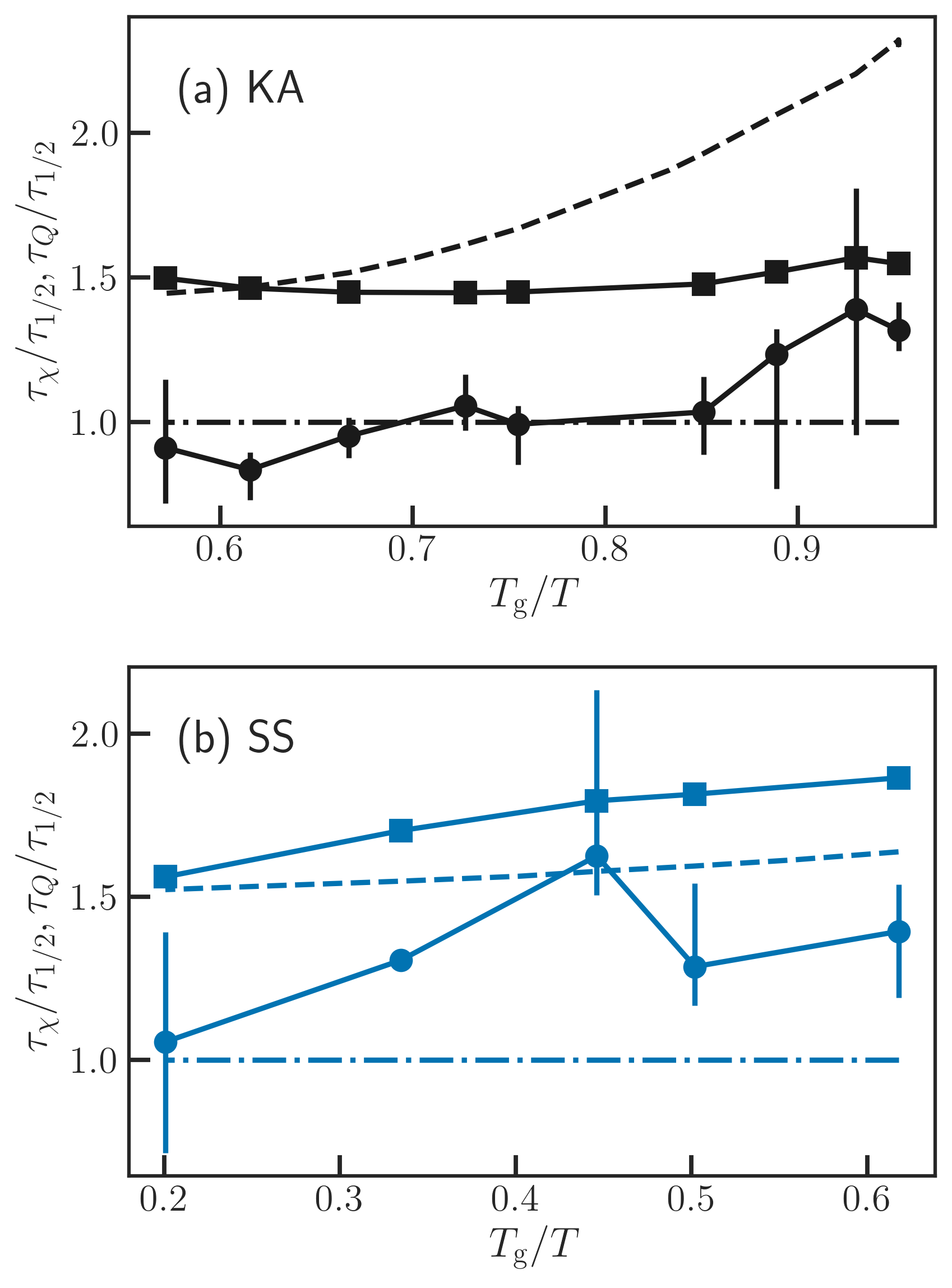}
\caption{ Ratios $\tau_\chi / \tau_{1/2}$ (circles) and $\tau_Q / \tau_{1/2}$ (squares), for (a) KA and (b) SS, $\rho=0.65$. Dashed lines are trap predictions. The comparison between these predictions and MD data is discussed in the supplementary text.}\label{fig:ratios}
\end{figure}

We have already seen in the main text that the trap-like model overestimates the stretching of $Q{\left(t\right)}$ in the KA system and underestimates it in the SS systems. Figure \ref{fig:ratios} elaborates on these facts by showing $\tau_Q / \tau_{1/2}$ and $\tau_\chi / \tau_{1/2}$ as a function of inverse temperature.  In the KA system we see that (1) at moderate supercooling, $\tau_\chi = \tau_{1/2}$ holds, and (2) indeed, $\tau_Q / \tau_{1/2}$, related to the degree of stretching of $Q$, grows much less with cooling than the model predicts. On the other hand, in the SS system we see that $\tau_\chi = \tau_{1/2}$ essentially never holds, and that $\tau_Q / \tau_{1/2}$ is larger, and grows faster with cooling, than predicted by the trap-like model. These facts are consistent with the picture discussed based on the stretching of $Q{\left(t\right)}$ at a single temperature in the main text.

\end{document}